\documentclass[preprint,12pt]{elsarticle}
\usepackage{amssymb}
\biboptions{sort&compress}

\begin{document}
%\begin{linenumbers}
\begin{frontmatter}



\title{Covariance and Causality in the Transition Radiation of an Electron Bunch}

\author{Gian Luca Orlandi}
\ead{gianluca.orlandi@psi.ch}
\address{Paul Scherrer Institut, 5232 Villigen PSI, CH.}

\begin{abstract}
A theoretical model of the transition radiation (TR) emission of an
$N$ electron bunch must comply with the covariance and the
temporal-causality principles. A charge-density-like covariance must
indeed imprint the formal expression of the TR energy spectrum. A
causality relation must constrain the emission phases of the
radiation pulse to the temporal sequence of the $N$ electron
collisions onto the metallic screen. Covariance and causality are
the two faces of the same coin: failing in implementing one of the
two constraints into the model necessarily implies betraying the
other one. The main formal aspects of a covariance and causality
consistent formulation of the TR energy spectrum of an N electron
beam will be here described with reference to the case of a radiator
surface with an arbitrary size.
\end{abstract}

\end{frontmatter}

%\linenumbers

\section{Introduction}

Transition radiation (TR) can be observed when a relativistic charge
is crossing a dielectric interface, for instance, a vacuum-metal
interface \cite{gifr,gari,gari2,frank2,bass,frank,ter,ginz}. The
dipolar oscillation of the conduction electrons induced on the
metallic surface by the incident relativistic charge is responsible
for the radiation emission. TR develops backward and forward from
the metallic surface with a characteristic angular distribution
scaling down with the inverse of the Lorentz $\gamma$ factor of the
charge ($\gamma=E/mc^2$). The well-known Ginzburg-Frank formula
accounts for the spectral and the angular distribution of the TR of
a single electron colliding onto a flat metallic screen whose size
is supposed to be infinite compared to the transverse extension of
the electromagnetic field traveling with the electron.

The present paper deals with the case of a bunch of $N$ electrons
colliding at a normal angle of incidence onto a flat metallic
surface, which is supposed to be placed in vacuum in the plane
$z=0$. The $N$ electrons are supposed to travel along the $z$-axis
of the laboratory reference frame with a rectilinear and uniform
motion and a common velocity $\vec{w}=(0,0,w)$. The radiation field
is supposed to be observed at a given point of the $z$-axis. With
reference to such a charge collision scenario, the roles played by
the longitudinal and transverse coordinates of the $N$ electrons in
determining the formal expression of the TR energy spectrum are
different in relation to the covariance and the temporal-causality
principles. The distribution of the $N$ electron longitudinal
coordinates, defining indeed the temporal sequence of the $N$
electron collisions onto the metallic screen, determines the
causality character of the emission phases of the radiation pulse
from the radiator surface. The distribution of the $N$ electron
transverse coordinates plays as well a role in determining the phase
of the radiation pulse at the observation point. In fact, as a
function of the distance of the given single electron of the bunch
from the $z$-axis of the laboratory reference frame, a further phase
delay depending on the transverse coordinate of the given electron
adds up, at the observation point, to the related emission phase of
the single electron contribution to the radiation pulse. In the
formal expression of the TR energy spectrum, the role of the $N$
electron transverse coordinates goes beyond the simple contribution
to the phase factor distribution of the radiation field at the
observation point \cite{giancov,gian-finite}. The distribution of
the $N$ electron transverse coordinates is indeed an invariant under
a Lorentz transformation with respect to the direction of motion of
the electron bunch. The signature of such a Lorentz invariance
intrinsically affects the $N$ single electron amplitudes composing
the radiation field. A covariant formulation of the TR energy
spectrum of a $N$ electron bunch is expected to preserve the
signature of such a Lorentz invariance characterizing the radiation
field in both the temporal coherent and incoherent components of the
spectrum \cite{giancov,gian-finite}.

In conclusion, passing from a single electron to a $N$ electron
bunch, the formal expression of the TR energy spectrum is expected
to show a charge-density-like covariance and to be causality
consistent \cite{giancov,gian-finite}. The failure in implementing
the causality in the formula of the TR energy spectrum necessarily
implies the failure of the covariance and viceversa. This will be
demonstrated in the following with reference to the general case of
a round radiator with an arbitrary radius \cite{gian-finite}. For
simplicity, ideal conductor properties for the metallic surface will
be supposed in the following.

\section{Transition Radiation Energy Spectrum}\label{pseudo-virtual}

Under far-field observation conditions, with reference to the charge
collision scenario considered in the present paper, the harmonic
component of the TR field of a $N$ electron bunch reads, see
\cite{giancov,gian,gianbis}:
\begin{eqnarray}
E_{x,y}^{tr}(\vec{\kappa},\omega)=
\sum_{j=1}^{N}H_{x,y}(\vec{\kappa},\omega,\vec{\rho}_{0j})\,e^{-i(\omega/w)z_{0j}},\label{tre}
\end{eqnarray}
where
\begin{eqnarray}
H_{\mu}(\vec{\kappa},\omega,\vec{\rho}_{0j})=H_{\mu,j}=\frac{iek}{2\pi^2Dw}\int\limits_S
d\vec{\rho}\int d\vec{\tau}
\frac{\tau_{\mu}e^{-i\vec{\tau}\cdot\vec{\rho}_{0j}}}{\tau^2+\alpha^2}
e^{i(\vec{\tau}-\vec{\kappa})\cdot\vec{\rho}}\label{quattro1}
\end{eqnarray}
with $\mu=x,y$. In previous equation,
$\vec{\rho}_{0j}=(x_{0j},y_{0j})$ and $z_{0j}$ $(j=1,..,N)$ are,
respectively, the transverse and the longitudinal coordinates of the
$N$ electrons in the laboratory reference frame at the time $t=0$
when the center of mass of the electron bunch is supposed to strike
the metallic surface; $D$ is the distance from the radiator surface
to the observation point which, in the present context, is supposed
to be on the $z$-axis of the laboratory reference frame;
$\vec{\rho}=(x,y)$ are the spatial coordinates of the radiator
surface $S$ which, in general, has an arbitrary shape and size
(either infinite $S=\infty$ or finite $S<\infty$); $k=\omega/c$ is
the radiation wave-number and $\vec{\kappa}=(k_x,k_y)$ is the
transverse component of the related wave-vector; finally,
$\alpha=\frac{\omega}{w\gamma}$ where $\vec{w}=(0,0,w)$ is the
common velocity and $\gamma$ the Lorentz factor of the electrons.

With reference to Eqs.(\ref{tre},\ref{quattro1}), the TR energy
spectrum of a $N$ electron beam is obtainable as the flux of the
Poynting vector:
\begin{eqnarray}
\frac{d^2I}{d\Omega
d\omega}&&=\frac{cD^2}{4\pi^2}\left(\left|E_{x}^{tr}(k_x,k_y,\omega)
\right|^2+\left|E_{y}^{tr}(k_x,k_y,\omega)\right|^2\right)=\label{cento-uno}\\
\nonumber
&&=\frac{cD^2}{4\pi^2}\sum_{\mu=x,y}\left(\sum_{j=1}^{N}\left|H_{\mu,j}
\right|^2+\sum_{j,l(j\neq l)=1}^{N}e^{-i(\omega/w)(z_{0j}-z_{0l})}
H_{\mu,j}H^*_{\mu,l}\right).
\end{eqnarray}
The size and the shape of the radiator surface $S$ being arbitrary
in Eqs.(\ref{tre},\ref{quattro1}), Eq.(\ref{cento-uno}) only states
the TR energy spectrum in an implicit form. As already argued in
\cite{giancov}, such an implicit formulation of the TR energy
spectrum meets the covariance and the temporal causality
constraints. The phase structure of the $N$ single electron
amplitudes composing the radiation field - see
Eqs.(\ref{tre},\ref{quattro1}) - is indeed causality related to the
temporal sequence of the $N$ electron collision onto the metallic
screen. A causality constraint characterizes the reciprocal
interference of the single electron radiation field amplitudes in
Eq.(\ref{cento-uno}) as well. About the covariance consistency of
the formulation of the TR energy spectrum as given in
Eqs.(\ref{tre},\ref{quattro1},\ref{cento-uno}), it can be
demonstrated \cite{giancov,gian-finite} that: (1) under a Lorentz
transformation from the laboratory to the rest reference frame, the
dependence of the charge electric field on the $N$ electron
transverse coordinates is a Lorentz invariant; (2) the Lorentz
invariant dependence of the charge electric field on the $N$
electron transverse coordinates transfers into the TR field leaving,
on both the temporal incoherent and coherent components of the TR
energy spectrum, a covariant imprinting whose observability is, in
principle, a function of the Lorentz invariant itself.

In the following, it will be demonstrated how the signature of the
causality and covariance principles may imprint the formula of the
TR energy spectrum or be vanished in Eq.(\ref{cento-uno}) depending
on the way how - in the expression of the radiation field
Eqs.(\ref{tre},\ref{quattro1}) - the integral calculus with respect
to the radiator surface $S$ is performed and, in particular, how the
limit $S\rightarrow\infty$ is implemented.

\subsection{Single Electron Ginzburg-Frank Formula}

The well-known Ginzburg-Frank formula accounts for the TR energy
spectrum of a single electron colliding onto a radiator surface
whose transverse size is much larger - at the limit, infinite -
compared to the transverse extension of the electromagnetic field
traveling with the relativistic electron. Under the limit
$S\rightarrow\infty$, in the case of a single electron $N=1$ - and
$\vec{\rho}_{0j}=(0,0)$ - the integral with respect to the spatial
coordinates $\vec{\rho}=(x,y)$ of the radiator surface $S$
transforms into a Dirac delta function in the formal expression of
the radiation field, see Eqs.(\ref{tre},\ref{quattro1}). This makes
easy the derivation of the Ginzburg-Frank formula via
Eq.(\ref{cento-uno}):
\begin{eqnarray}
\frac{d^2I_{e}}{d\Omega
d\omega}=\frac{(e\beta)^2}{\pi^2c}\frac{sin^2\theta}
{(1-\beta^2cos^2\theta)^2}.\label{nove}
\end{eqnarray}
The single electron Ginzburg-Frank formula can be obtained from
Eqs.(\ref{tre},\ref{quattro1},\ref{cento-uno}) by applying the limit
$S\rightarrow\infty$, equivalently, either before performing the
integral calculus of the radiation field with respect to the
radiator surface - see above - or after performing the integral
calculus with respect to a screen surface with a finite size
($S<\infty$), see
\cite{shulga1,poty,castellano,casalbuoni1,sutter,casalbuoni2}. In
order to calculate the TR energy spectrum of a single electron
colliding onto an infinite metallic surface, the two mathematical
procedures are equivalent and lead to the same result, see
Eq.(\ref{nove}).

\subsection{N Electron Bunch Formula}

Contrary to the case of a single electron, in the case of a $N$
electron bunch the above mentioned two mathematical procedures to
implement the limit $S\rightarrow\infty$ in
Eqs.(\ref{tre},\ref{quattro1}) - either before or after the integral
calculus - lead to two completely different results, as in the
following described.

In the case of a $N$ electron bunch, if the limit to infinity of the
metallic surface ($S\rightarrow\infty$) is performed in
Eqs.(\ref{tre},\ref{quattro1}) prior to the integral calculus of the
radiation field, then the following expression of the TR field can
be obtained:
\begin{eqnarray}
E_{x,y}^{tr}(\vec{\kappa},\omega)=
\sum_{j=1}^{N}E_{x,y}^{e}(\vec{\kappa},\omega)\,e^{-i(\omega/w)z_{0j}}\,e^{-i\vec{\kappa}\cdot\vec{\rho}_{0j}},\label{eins}
\end{eqnarray}
where
\begin{eqnarray}
E_{x,y}^{e}(\vec{\kappa},\omega)=\frac{2iek}{Dw}\frac{\kappa_{x,y}}{\kappa^2+\alpha^2}\label{zwei}
\end{eqnarray}
is the harmonic component of the TR field produced by a single
electron moving along the $z$-axis of the laboratory reference frame
and
$\vec{\kappa}=(\kappa_x,\kappa_y)=k\sin\theta(\cos\phi,\sin\phi)$ is
the transverse component of the wave-vector ($k=2\pi/\lambda$). With
reference to Eq.(\ref{cento-uno}), taking into account
Eqs.(\ref{eins},\ref{zwei}), the TR energy spectrum of a $N$
electron bunch results to be described by the following formula:
\begin{eqnarray}
\frac{d^2I}{d\Omega d\omega}&&=\frac{d^2I_e}{d\Omega
d\omega}\left(N+\sum_{j,l(j\neq
l)=1}^{N}e^{-i(\omega/w)(z_{0j}-z_{0l})}\,e^{-i\vec{\kappa}\cdot(\vec{\rho}_{0j}-\vec{\rho}_{0l})}\right)\label{drei}
\end{eqnarray}
where $\frac{d^2I_e}{d\Omega d\omega}$ is the Ginzburg-Frank formula
of a single electron, see Eq.(\ref{nove}), and the double summation
is proportional to the charge factor which, under the continuous
limit approximation, reads as the square module of the Fourier
transformation of the distribution function of the $N$ electron
spatial coordinates.

Conversely, if the integral calculus in
Eqs.(\ref{tre},\ref{quattro1}) is performed with respect to a
metallic screen with a finite size - a round metallic screen with a
finite radius $R$ $(R\gg\rho_{0j}, j=1,..,N)$, for instance - then
the formula of the radiation field of a $N$ electron bunch reads,
see \cite{gian-finite}:
\begin{eqnarray}
E_{x,y}^{tr}(\vec{\kappa},\omega)&&=\sum_{j=1}^{N}H_{x,y}(\vec{\kappa},\omega,\vec{\rho}_{0j})\,e^{-i(\omega/w)z_{0j}}=
\sum_{j=1}^{N}\frac{2iek}{Dw}\frac{\kappa}{\kappa^2+\alpha^2}
\,e^{-i[(\omega/w)z_{0j}+\vec{\kappa}\cdot\vec{\rho}_{0j}]}\times\nonumber\\\label{vier}
&&\times\left(\begin{array}{c}
\cos\phi\\
\sin\phi\\
\end{array}\right)[\rho_{0j}\Phi(\kappa,\alpha,\rho_{0j})-(R+\rho_{0j})\Phi(\kappa,\alpha,R+\rho_{0j})],
\end{eqnarray}
where
$\vec{\kappa}=(\kappa_x,\kappa_y)=k\sin\theta(\cos\phi,\sin\phi)$ is
the transverse component of the wave-vector with $k=2\pi/\lambda$
and $\vec{\rho}_{0j}=(x_{0j},y_{0j})$ $(j=1,..,N)$ are the
transverse coordinates of the $N$ electrons with
$\rho_{0j}=\sqrt{x_{0j}^2+y_{0j}^2}$. Furthermore,
\begin{eqnarray}
\Phi(\kappa,\alpha,\rho_{0j})=\alpha
J_0(\kappa\rho_{0j})K_1(\alpha\rho_{0j})+\frac{\alpha^2}{\kappa}
J_1(\kappa\rho_{0j})K_0(\alpha\rho_{0j})\label{funf}
\end{eqnarray}
and
\begin{eqnarray}
\Phi(\kappa,\alpha,R+\rho_{0j})=\alpha
J_0[\kappa(R+\rho_{0j})]K_1[\alpha(R+\rho_{0j})]+\frac{\alpha^2}{\kappa}
J_1[\kappa(R+\rho_{0j})]K_0[\alpha(R+\rho_{0j})].\label{sechs}
\end{eqnarray}
It can be demonstrated \cite{gian-finite} that, in the limit
$R\rightarrow\infty$,
\begin{eqnarray}
(R+\rho_{0j})\Phi(\kappa,\alpha,R+\rho_{0j})\rightarrow0.\label{sieben}
\end{eqnarray}
Moreover, in the case $N=1$ and under the limit
$\rho_{01}\rightarrow0$, it can be demonstrated \cite{gian-finite}
that the model represented by
Eqs.(\ref{vier},\ref{funf},\ref{sechs},\ref{sieben}) can reproduce
the well known formula of the TR energy spectrum of a single
electron colliding onto a round metallic screen with a finite radius
$R$ \cite{shulga1,poty,castellano,casalbuoni1,sutter,casalbuoni2}.
See, for instance, the comparison of the results reported in
\cite{gian-finite} and in \cite{casalbuoni2}. Furthermore, it can be
demonstrated \cite{gian-finite} that the Ginzburg-Frank formula -
Eq.(\ref{nove}) - can be obtained from
Eqs.(\ref{vier},\ref{funf},\ref{sechs},\ref{sieben}) under the
limits $\rho_{01}\rightarrow0$ and $R\rightarrow\infty$.

\begin{figure*}[tb]
   \centering
   \includegraphics*[width=80mm]{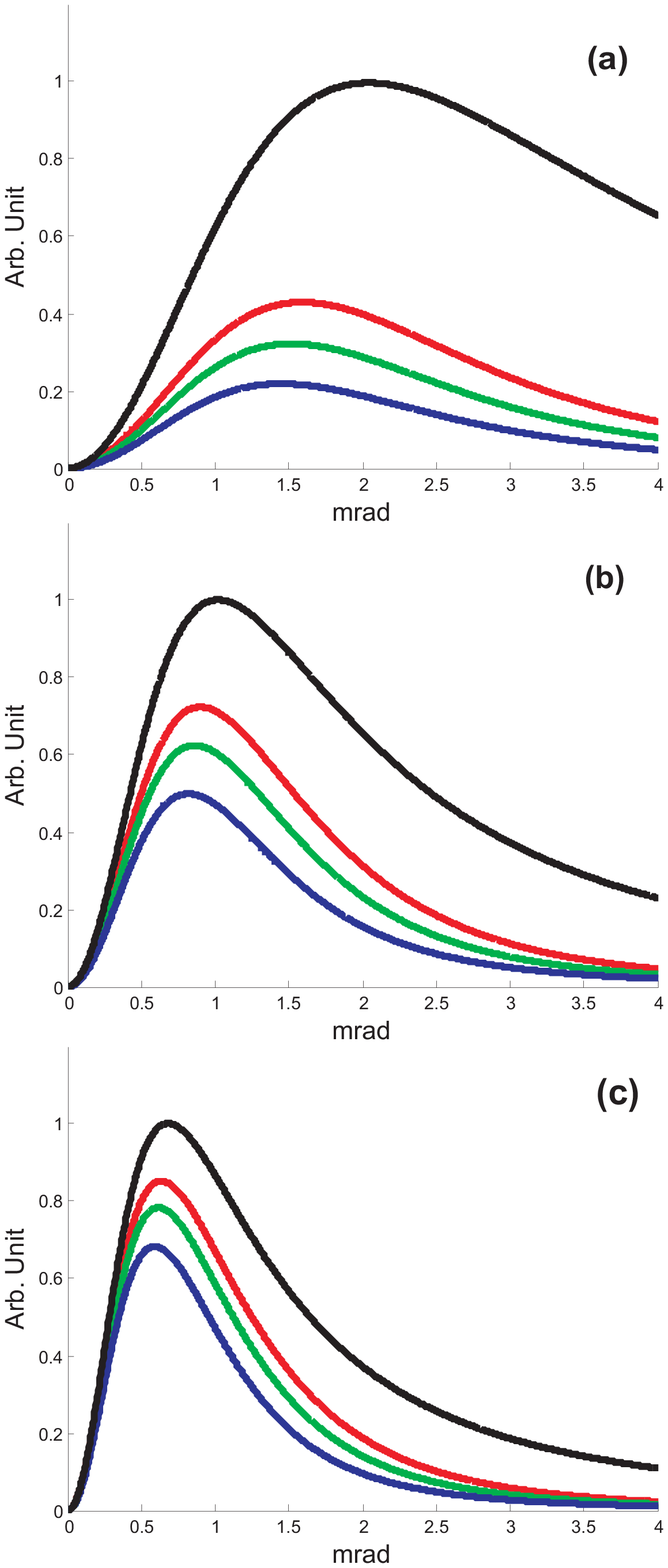}
   \vspace*{-\baselineskip}
   \caption{Numerical simulation of the angular distribution of the TR originated
   by a bunch of $N=10^4$ electrons with a gaussian transverse distribution of standard deviation
   $\sigma=50$ $\mu m$. Different beam energy are supposed: $250$ $MeV$ (a), $500$ $MeV$ (b) and $750$ $MeV$ (c).
   The Red, Green, and Blue curves are the numerical results that are obtainable from the first
   term of Eq.(\ref{acht}) with reference to Eqs.(\ref{nove},\ref{funf},\ref{neun}) for a wavelength
   $\lambda=680$ $nm$, $\lambda=530$ $nm$ and $\lambda=400$ $nm$, respectively.
   The Black curves represent the numerical results obtainable from
   the first term of Eq.(\ref{drei}).}
   \label{Fig1}
\end{figure*}

Finally, according to the model represented by
Eqs.(\ref{vier},\ref{funf},\ref{sechs},\ref{sieben}), under the
limit $R\rightarrow\infty$, the TR energy spectrum of a $N$ electron
bunch colliding onto a radiator surface with an infinite size
($S=\infty$) can be finally formulated via Eq.(\ref{cento-uno}) as,
see also \cite{gian-finite}:
\begin{eqnarray}
\frac{d^2I}{d\Omega d\omega}=\frac{d^2I_{e}}{d\Omega
d\omega}\left(\sum_{j=1}^{N}|A_j|^2+\sum_{j,l (j\neq
l)=1}^{N}A_jA^*_l\,e^{-i[(\omega/w)(z_{0j}-z_{0l})+\vec{\kappa}\cdot(\vec{\rho}_{0j}-\vec{\rho}_{0l})]}\right)\label{acht}
\end{eqnarray}
where $\frac{d^2I_{e}}{d\Omega d\omega}$ is the Ginzburg-Frank
formula, see Eq.(\ref{nove}), and
\begin{eqnarray}
A_j=\rho_{0j}\Phi(\kappa,\alpha,\rho_{0j}).\label{neun}
\end{eqnarray}
More details on the derivation of the formulae above in
\cite{gian-finite}. Numerical results of the angular distribution of
the spectral intensity of the TR, which are obtainable from
Eq.(\ref{acht}) and Eq.(\ref{drei}) under an observation condition
of temporal incoherence, are compared in Fig.(\ref{Fig1}). In
Fig.(\ref{Fig1}), it can be observed that, for given values of the
beam energy and size, the angular distribution of the TR undergoes a
broadening with the increase of the observed wavelength. Such an
angular broadening of the TR spectral intensity is consistent with a
diffractive effect due to the finite transverse size of the electron
beam in comparison with the observed wavelength.

Contrary to the case of a single electron, in the case of an $N$
electron bunch different formulae of the TR energy spectrum - see
Eq.(\ref{drei},\ref{acht}) - follows from the two different
mathematical methods to implement the limit to infinity of the
metallic surface ($S\rightarrow\infty$) in the integral calculus of
the radiation field, see Eqs.(\ref{tre},\ref{quattro1}).

If the limit $S\rightarrow\infty$ is implemented in
Eqs.(\ref{tre},\ref{quattro1}) before performing the integral
calculus of the TR field - see Eq.(\ref{drei}) - then the causality
role played by the longitudinal coordinates of the $N$ electrons in
determining the emission phases of the $N$ single electron
amplitudes composing the radiation field becomes indistinguishable
from the role played by the $N$ electron transverse coordinates
which, in principle, are only expected to contribute to the relative
phase distribution of the radiation field at the observation point
with an additional phase delay adding up to the emission phase. In
practice, looking at Eq.(\ref{eins}) and Eq.(\ref{drei}), the
temporal causality feature characterizing the radiation emission
cannot be univocally attributed to the distribution function of the
$N$ electron longitudinal coordinates $z_{0j}$ $(j=1,..,N)$.
Moreover, the expected Lorentz invariant signature of the $N$
electron transverse coordinates on the radiation field is completely
lost since, for a possible observer of the radiative mechanism under
an observation condition of temporal incoherence, the radiation
field results to be the linear addition of $N$ identical single
electron radiation field contributions originated by an electron
traveling exactly on the $z$-axis of the laboratory reference frame.

On the contrary, if the limit $S\rightarrow\infty$ is implemented in
Eqs.(\ref{tre},\ref{quattro1}) after performing the integral
calculus of the TR field then - in the formula of the TR energy
spectrum, see Eq.(\ref{acht}) - the causality role played by the $N$
electron longitudinal coordinates in determining the emission phases
of the $N$ single electron radiation field amplitudes maintains
distinct from the role played by the $N$ electron transverse
coordinates which are only expected to contribute with a further
phase delay to the observation phase distribution of the radiation
field. Moreover, both the temporal coherent and incoherent
components of the TR energy spectrum - see Eq.(\ref{acht}) - bear
the signature of the Lorentz invariance of the distribution function
of the $N$ electron transverse coordinates. The $N$ single electron
radiation field amplitudes - see Eq.(\ref{vier}) - being indeed an
intrinsic function of such a Lorentz invariant quantity characterize
with a charge-density-like covariance the formula of the TR energy
spectrum, see Eq.(\ref{acht}).

\section{Conclusions}

In the collision of a $N$ electron bunch at a normal angle of
incidence onto a flat metallic surface with an arbitrary size, the
longitudinal and the transverse coordinates of the $N$ electrons
bring, respectively, into the formal expression of the TR energy
spectrum the causality and the covariance marks characterizing the
electromagnetic radiative mechanism. In the case of a $N$ electron
bunch, in the present paper it is demonstrated how an improper
mathematical procedure to implement the limit of infinite surface in
the integral calculus of the radiation field can lead to the
non-physical result of mixing up and, in conclusion, losing the
distinct roles which the longitudinal and transverse coordinates of
the $N$ electrons play, respectively, in determining the causality
and the covariance features of the TR emission.

%\section{ACKNOWLEDGMENTS}

%\section{}
%\label{}

% \appendix

% \section{}
% \label{}

%\end{linenumbers}

\end{document}